\documentclass[12pt]{article}
\usepackage[utf8]{inputenc}
\usepackage[english]{babel}
\usepackage{amsmath}
\usepackage{amsfonts}
\usepackage{amssymb}
\usepackage{graphicx}
\usepackage{pgf,tikz}
\usepackage{mathrsfs}
\usetikzlibrary{arrows}

\def\C{\mathbb C}
\def\R{\mathbb R}

\def\r{\rangle}
\def\l{\langle}

\def\o{\omega}
\def\l{\langle}
\def\r{\rangle}
\def\hs{\hspace{0.5 cm}}

\newcommand{\mbra}[1]{\langle {#1}|}
        \newcommand{\mket}[1]{ |{#1}\rangle}

        \newcommand{\mop}[1]{\hat{#1}}

\author{G.~Chadzitaskos
\footnote{Faculty of Nuclear Sciences and Physical Engineering,
Czech Technical University in Prague, B\v rehov\'a 7, CZ - 115 19 Prague, Czech Republic, e-mail: goce.chadzitaskos@fjfi.cvut.cz}, J.~Patera\footnote{October 10, 1936, Zdice - January 3, 2022, Montreal ( Centre de Recherches Mathematiques, Universit\' e  de Montreal, C.P. 6128 Succ. Centre-ville, Montreal, Qc, H3C 3J7, Canada) }
 } 
\title{Asymmetric harmonic oscillator}

\begin{document}

\maketitle
\abstract{The solution of one--dimensional asymmetric quantum harmonic oscillator is presented. The asymmetry can be realized, for example, by using two springs, one spring is glued with the mass, and the second spring is freely connected with the mass in the equilibrium point and it is located inside or outside the first spring which acts on the mass only from the contact point on the right. 
We study the spectrum of a quantum harmonic oscillator, which has a spring constant $ k _- $ to the left of the equilibrium position and a spring constant $ k _ +$ to the right of the equilibrium position.
In the presented case the contact point of the second string is the equilibrium point of the first string. The explicit form of eigenfunctions, the way to calculate the eigenvalues and the properties of the eigenfunctions are discussed.} 


\section{Introduction}

 We present solution of one--dimensional harmonic oscillator under two spring forces. One for $x\geq 0$ with characteristic frequency $\omega_+ = \sqrt{k_+/m}$ and the other for $x<0$ with characteristic frequency $\omega_- =\sqrt{k_-/m}.$  Such oscillator can be realized as a mass fitted on a spring, with second spring outside or inside the first one not connecting the mass. The Van der Waals forces acting on oscillating atoms on the surface can lead to a similar effect. The interesting feature of such an oscillator is that the energy levels between neighboring states are not equidistant in general. To pass from one state to another state it is necessary to emit or absorb a photon of specific frequencies.

\section{Classical case}

 We study the simple case postulating that the equilibrium position of the mass is the point where the forces of both strings are zero.
 
  The classical harmonic oscillator with different spring forces is described by the equation of motion
  $$\frac{d^2 x }{d t^2} + \omega^2(x)x=0,\hs \o (x)= \lbrace \begin{smallmatrix}\o_+ \hspace{.2 cm} \mbox{for} \hspace{.1cm} x<0\\  \o_- \hspace{.2cm} \mbox{for} \hspace{.1 cm} x\geq 0 \end{smallmatrix},$$
with the Hamiltonian
\begin{equation}
{\cal H}= \frac{p_{x}^{2}}{2m} + \frac{1}{2}m \omega^{2}x^{2}.
\end{equation}
The solution is a periodic motion with the period $T$
$$ T=\frac{ \pi}{\o_-} + \frac{\pi}{\o_+}.$$ 
 The energy conservation guarantees that the velocities are continuous functions of the time, see Fig.1.
 \begin{figure}[h]
 \caption{Graph of motion of a classical asymmetric oscillator}
 \includegraphics[scale=0.4]{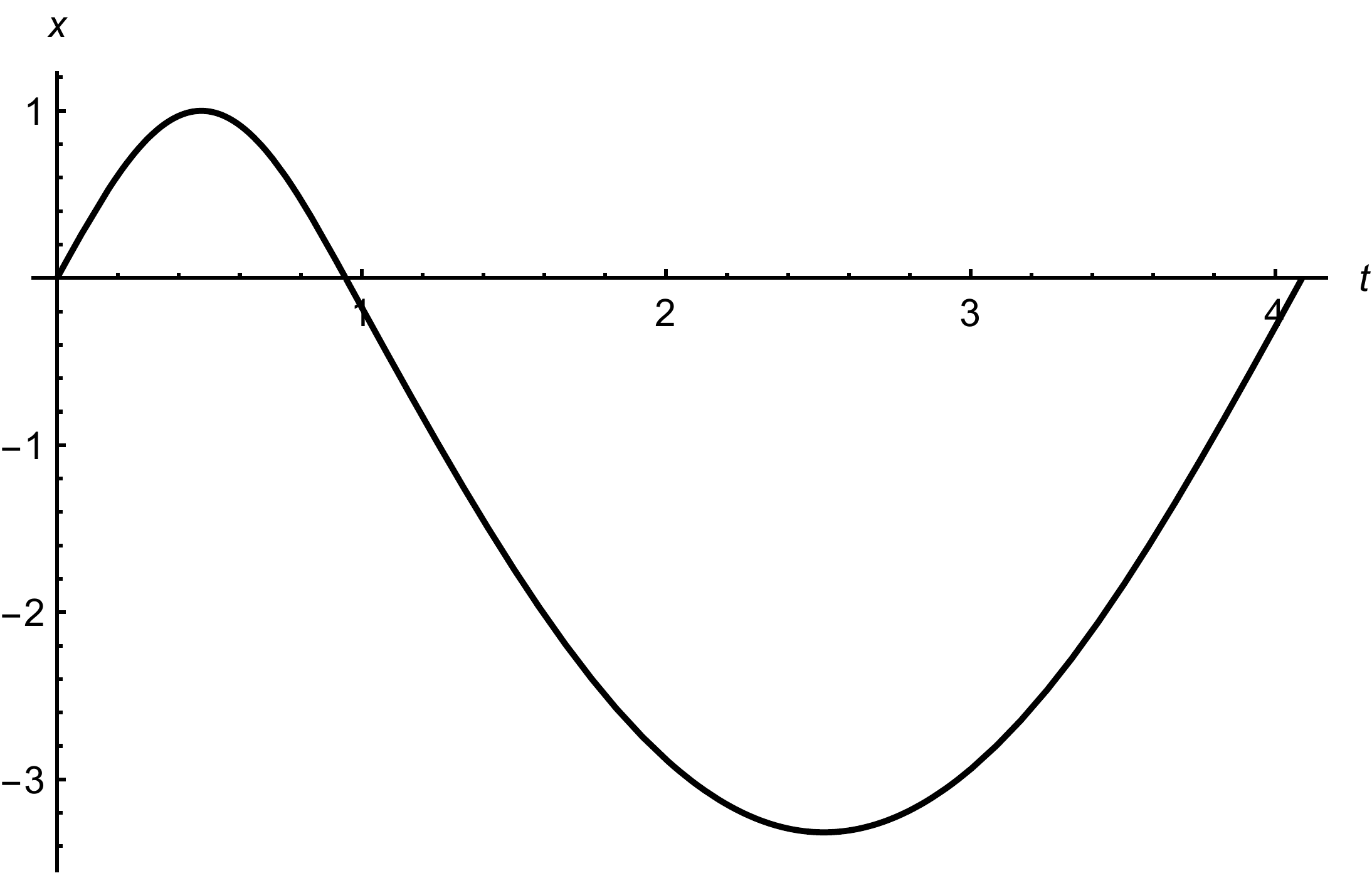}
 \end{figure}

 \section{Quantum case}

 Let the Hamilton operator be the sum of the two separate Hamilton operators one on the positive half-line and the second on the negative half-line.
 
 \begin{equation}{\label{AsH}}
\hat H=\hat H_-+ \hat H_+ = \frac{\hat P_{x}^{2}}{2m} + \frac{1}{2}m \omega(x)^{2}\hat{X}^{2} =  \lbrace \begin{smallmatrix} - \frac{\hbar^2}{2m}\frac{d^2}{d x^2}+\frac{1}{2}m \omega_+^{2}x^{2}, \hspace{.1cm} \mbox{for} \hspace{.1cm} x \geq 0\\ - \frac{\hbar^2}{2m}\frac{d^2}{d x^2}+\frac{1}{2}m \omega_-^{2} x^{2}, \hspace{.1cm} \mbox{for} \hspace{.1cm}  x< 0 \end{smallmatrix}
\end{equation}
The Schr\"odinger equation 
\begin{equation}{\label{Sch}}
 \hat H \psi (x) = E\psi (x)
\end{equation}
is the equation we have to solve. 

\subsection{Symmetric harmonic oscillator}

For the ordinary symmetric harmonic oscillator (the case  $\omega_+ = \omega_- =\omega$) the solutions of the Schr\" odinger equation are given by tha eigenfunctions
 \begin{equation}
{\psi_{n}(x)}=\frac{1}{\sqrt{ 2^{n} n!}}  (\frac{m \omega}{\pi
\hbar})^{1/4} \exp[-m \omega x^2 /2\hbar ] H_{n} (\sqrt{\frac{m \omega}{\hbar}} x),\,\,n=0,1,\ldots,
\end{equation}
where $H_n$ are Hermite polynomials
$$H_n(x)= \sum_{k=0}^{[n/2]} \frac{(-1)^k n!}{k! (n-2k)} (2x)^{n-2k}, $$
and the corresponding eigenvalues are
$$E_n = \hbar \omega (n + \frac{1}{2}).$$

The symmetry $\omega_+ = \omega_-$ causes significant simplification when solving this problem.

\subsection{Asymmetric harmonic oscillator}

To solve  the  Schr\"odinger equation (\ref{Sch})  for the asymmetric harmonic oscillator (\ref{AsH}), i.e.  $\hat{H} \psi(x) = E \psi(x) $ on the negative and positive real half--axes 
\begin{equation}{\label{Sch}} 
[- \frac{\hbar^2}{2m}\frac{d^2}{d x^2}+\frac{1}{2}m \omega_\pm^{2} x^{2}] \psi(x)=E \psi(x),
\end{equation}
we substitute
\begin{equation}
\xi_\pm = \sqrt{\frac{2m\omega_\pm}{\hbar}} x, \,\,\, E=\hbar \omega_\pm(\nu_\pm +\frac{1}{2}), 
\end{equation}
and we get 
\begin{equation}
[\frac{d^2}{d \xi_\pm^2}-\frac{1}{4}\xi_\pm^{2} +\nu_\pm +\frac{1}{2}] \psi(\xi_\pm)=0.
\end{equation}
This is the Weber equation \cite{RG} (9.255) on positive and negative half-axes separately, where  $\xi_- \in ( \infty,0)$ and $\xi_+ \in \langle 0, \infty).$ 

Solutions of the Weber equation \cite{RG}(9.255) on $L^2(\R^+)$ are  parabolic cylinder functions $D_{\nu_+}(\xi)$ (for positive values of argument  $\xi$)  and they are  $D_{\nu_-}(-\xi)$ on $L^2(\R^-)$ ( for negative values of argument $\xi$). The eigenfunctions of the Hamilton oprator on the real axis consist of two different parts - on the negative half--axis and on the positive half--axis

$$
\psi(x)_{\nu}=\lbrace \begin{smallmatrix} \alpha_{\nu_-} \, D_{\nu_-}(-\sqrt{\frac{2m \omega_-}{\hbar}} x) ,  \mbox{ for }  x<0\\ \alpha_{\nu_+} \, D_{\nu_+}(\sqrt{\frac{2m \omega_+}{\hbar}} x), \hspace{.1cm} \mbox{ for } \hspace{.1cm}  x\geq 0 \end{smallmatrix}.
$$
Because  $\psi (x)_\nu $  has to be eigenfunction of operator $\hat{H}$ on the real axis with eigenvalue $E_\nu$, the $\nu_-$ can be expressed as a linear function of $\nu_+.$  For the relations between $\nu_+$ and $\nu_-$ we get
\begin{eqnarray}{\label{nupm}}
E_\nu=\hbar \omega_+(\nu_+ + \frac{1}{2})=\hbar \omega_-(\nu_- + \frac{1}{2}), \, \, s=\frac{\omega_+}{\omega_-},\,\, \nu_-=s \nu_+ + \frac{s-1}{2}.
\end{eqnarray}

Furthermore, we can assume $s \geq 1$ without loss of generality. The case $s=1$ is the case of a symmetric harmonic oscillator.

Continuity requirements of the eigenfunctions and their derivatives at $x=0$ give the following conditions
\begin{itemize}
\item $\alpha_{\nu_-} D_{\nu_-}(0) = \alpha_{\nu_+} D_{\nu_+} (0)$
\item $-\alpha_{\nu_-} \sqrt{\frac{2m \omega_-}{\hbar}}D'_{\nu_-} (0)  = \alpha_{\nu_+}\sqrt{\frac{2m \omega_+}{\hbar}}D'_{\nu_+} (0).$
\end{itemize} 

The first condition guarantees the continuity of eigenfunctions at $x=0,$ and the second condition guarantees the continuity of the derivatives of the eigenfunctions at $x=0.$
From the definition of parabolic cylinder functions $D_\nu (x)$ we get values at $x=0$
\begin{equation}
D_{\nu} (0)=\frac{\sqrt{\pi}\, 2^{\frac{\nu}{2}}}{\Gamma(\frac{1-\nu}{2})}, \,\,\,  D'_{\nu} (0)= - \frac{\sqrt{\pi}\, 2^{\frac{\nu +1}{2}}}{\Gamma(-\frac{\nu}{2})},
\end{equation}
and the previous conditions lead to the system of two linear equations
\begin{equation}
\alpha_{\nu_-}\frac{ 2^{\frac{\nu_-}{2}}}{\Gamma(\frac{1-\nu_-}{2})} - \alpha_{\nu_+}\frac{ 2^{\frac{\nu_+}{2}}}{\Gamma(\frac{1-\nu_+}{2})} = 0,\,\,\,
\alpha_{\nu_-} \frac{\sqrt{\omega_-}\, 2^{\frac{\nu_- +1}{2}}}{\Gamma(-\frac{\nu_-}{2})} + \alpha_{\nu_+}\frac{\sqrt{\omega_+}\, 2^{\frac{\nu_+ +1}{2}}}{\Gamma(-\frac{\nu_+}{2})} =0.
\end{equation}
This system of equations has nontrivial solutions if and only if 
\begin{equation}
\frac{1}{\Gamma(\frac{1-\nu_-}{2})} \frac{ 1}{\Gamma(-\frac{\nu_+}{2})} +
\frac{1}{\sqrt{s}} \frac{1}{\Gamma(-\frac{\nu_-}{2})} \frac{ 1}{\Gamma(\frac{1-\nu_+}{2})}  =0.
\end{equation}
Because of relations (\ref{nupm}) this equation determines the set $\{\nu_+^{(i)} |\, i= 1,2,\ldots  \},$ and the spectrum of the operator $\hat{H} $ is the set $\Omega\equiv \{E_i =\hbar \omega_+ (\nu_+^{(i)}+ \frac{1}{2}) |\, i= 0,1,2,\ldots   \}.$
 Then the set of eigenfunctions $\{\psi_i| E_i \in \Omega  \}$ is orthogonal basis in $L^2(\R).$ ( If we choose constants $\alpha_{\nu_+} $ and $\alpha_{\nu_-} $ so that the eigenfunctions are normalized, then the basis is orthonormal.)

 When $ D_{\nu_\pm}(0)\neq 0,$ we get from both equations
\begin{equation}{\label{Fv}}
\frac{D'_{\nu_-} (0)}{ D_{\nu_-}(0)}=-\sqrt{s}\frac{D'_{\nu_+} (0)}{ D_{\nu_+}(0)}
\end{equation}
The cases $D'_{\nu_\pm} (0) = 0$ and  $ D_{\nu_\pm}(0)=0$ lead to even and odd Hermite functions respectively. 
Let us introduce function
$$ F(\nu)=\frac{1}{\sqrt{2}}\frac{D'_{\nu} (0)}{ D_{\nu}(0)}= \frac{\Gamma(\frac{1-\nu}{2})}{\Gamma(-\frac{\nu}{2})},$$
then Eq. (\ref{Fv}) leads to equation
\begin{equation}
F(\nu_+) = - \frac{1}{\sqrt{s}} F (\nu_-)= - \frac{1}{\sqrt{s}} F (s\nu_+ + \frac{s-1}{2}),
\end{equation}
This is the crucial transcendental equation that must be solved in order to find the spectrum and the eigenfunctions of the Hamiltonian of the asymmetric harmonic oscillator.

\subsection{Properties of solutions and positions of asymptotes}
There are several properties of the function $F(\nu).$ They follow from the properties of gamma  and digamma functions.
\begin{itemize}
\item Function $F(\nu)$ is monotone decreasing in intervals $(-\infty,1)$ and $(2M+1,2M+3), M=0,1,\ldots$
\item Function $F(\nu)$ has zeros at points $2M, M=0,1,\ldots$
\item Function $F(\nu)$ has left limit $- \infty$ and right limit $+ \infty$  at asymptotic points $2M+1\,, M=0,1,\ldots.$ 
\end{itemize}
The direct consequence of the previous properties is the following statement: function $ - \frac{1}{\sqrt{s}} F (s\nu + \frac{s-1}{2})$ is monotone increasing in intervals $(-\infty,\frac{3}{2s}- \frac{1}{2})$ and $(\frac{4M+3}{2s}-\frac{1}{2},\frac{4M+7}{2s}-\frac{1}{2}), M=0,1,\ldots$ with left limit  $+ \infty$ and right limit  $- \infty$ at the limit points of the intervals, i. e. at the asymptotes. The distance between the asymptotes is 2 or  $\frac{2}{s},$ respectively.  See the example in Fig.2.

 \begin{figure}[h]{\label{Fv}}
\includegraphics[scale=0.4]{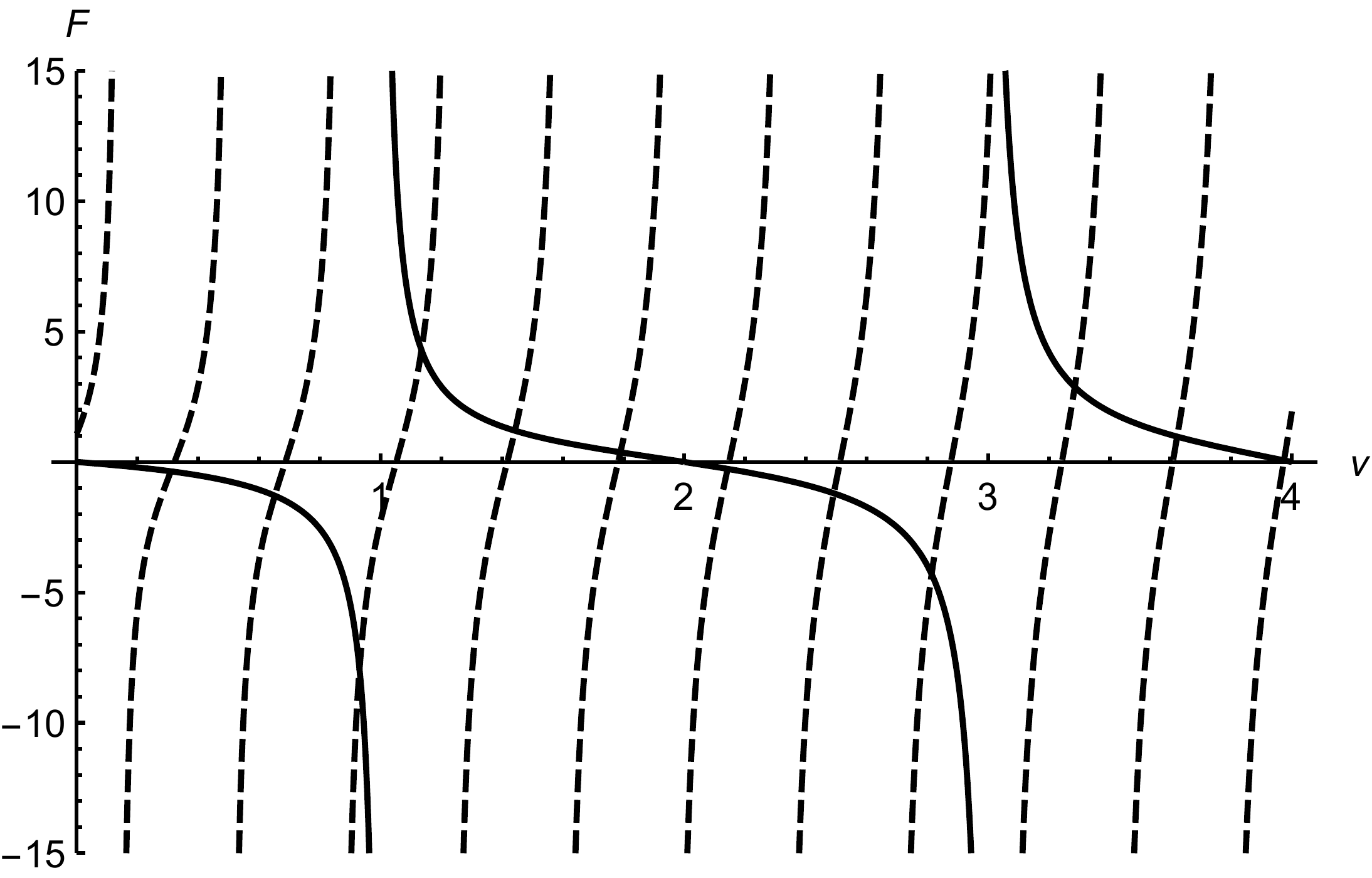}
\caption{Functions $F(\nu) = \frac{ \Gamma \left(\frac{1-\nu}{2}\right)}{\Gamma \left(-\frac{\nu}{2}\right)}$ - solid lines and $-F(\sqrt{30}\nu+\frac{\sqrt{30}-1}{2})$ - dashed line. The crossing points of the lines correspond to the values of $\nu=\nu_+.$  }
\end{figure}

 \subsection{Eigenvectors and eigenvalues} 
 The energy spectrum is $E_\nu = \hbar \omega_+(\nu_+ + \frac{1}{2}),$ where values $\nu_+$ correspond to cross points of solid and dashed lines on the graph on Fig. 2.
  
 {\bf Remarks:} 
 \begin{itemize}
 \item The eigenfunctions depend only on the ratio of the proper frequencies on positive and negative half lines (i.e. on the classical level -- on the ratio of square roots of spring constants in the positive and negative directions).
 \item The values have to be calculated numerically
 \item As the ratio of spring constants increases, the ground energy level goes to zero. 
 \item In the special cases where $s=\frac{\omega_+}{\omega_-}= \frac{4m+3}{4n+3},\,\, m,n=0,1,\ldots,$ some eigenfunctions are constructed by gluing two Hermite functions, one on negative and one on positive half-axis. The energies (eigenvalues) corresponding to these eigenfunctions are "equidistant".   
 \end{itemize}

 \begin{table}[h]
\caption{Examples of first 8 eigenvalues of $\nu_+$ for different parameter $s$}
\begin{small}
\begin{tabular}{|l|l|l|l|l|l|l|l|l|}

\hline
$s$&$\nu_{0}$&  $\nu_{1}$& $\nu_{2}$ &$\nu_{3}$& $\nu_{4}$&$\nu_{5}$&$\nu_{6}$&$\nu_{7}$ \\
\hline
1& 0&1&2&
3&4&5&6&
  7\\
\hline
1.4& -0.0815358& 0.748707&1.5841&
  2.41625&3.25019& 4.08329&4.91663&
  5.75007\\
\hline
$\sqrt{5}$ &-0.183585& 0.423418& 1.04532& 1.66393& 2.2807& 2.89906&3.51751&4.13516\\
\hline
$\sqrt{11}$ & -0.256549 & 0.192094&  0.656273&
  1.12213 &1.58579& 2.0482&2.51118& 
  2.97491\\
  \hline
4& -0.286739&0.0982773& 0.497364&
  0.899531&1.3008& 1.70056&2.09984&
  2.49961\\
  \hline 
5&-0.318944& 0& 0.330721& 0.665139&1&
  1.33404& 1.66719& 2.33301\\
  \hline
$\sqrt{30}$ &-0.330956& -0.0361063& 0.269545&
  0.578901& 0.889065& 1.19877&1.50768 &1.81608\\
  \hline
  \end{tabular}
  
\end{small}
\end{table}

\begin{figure}[h]
  \includegraphics[scale=0.4]{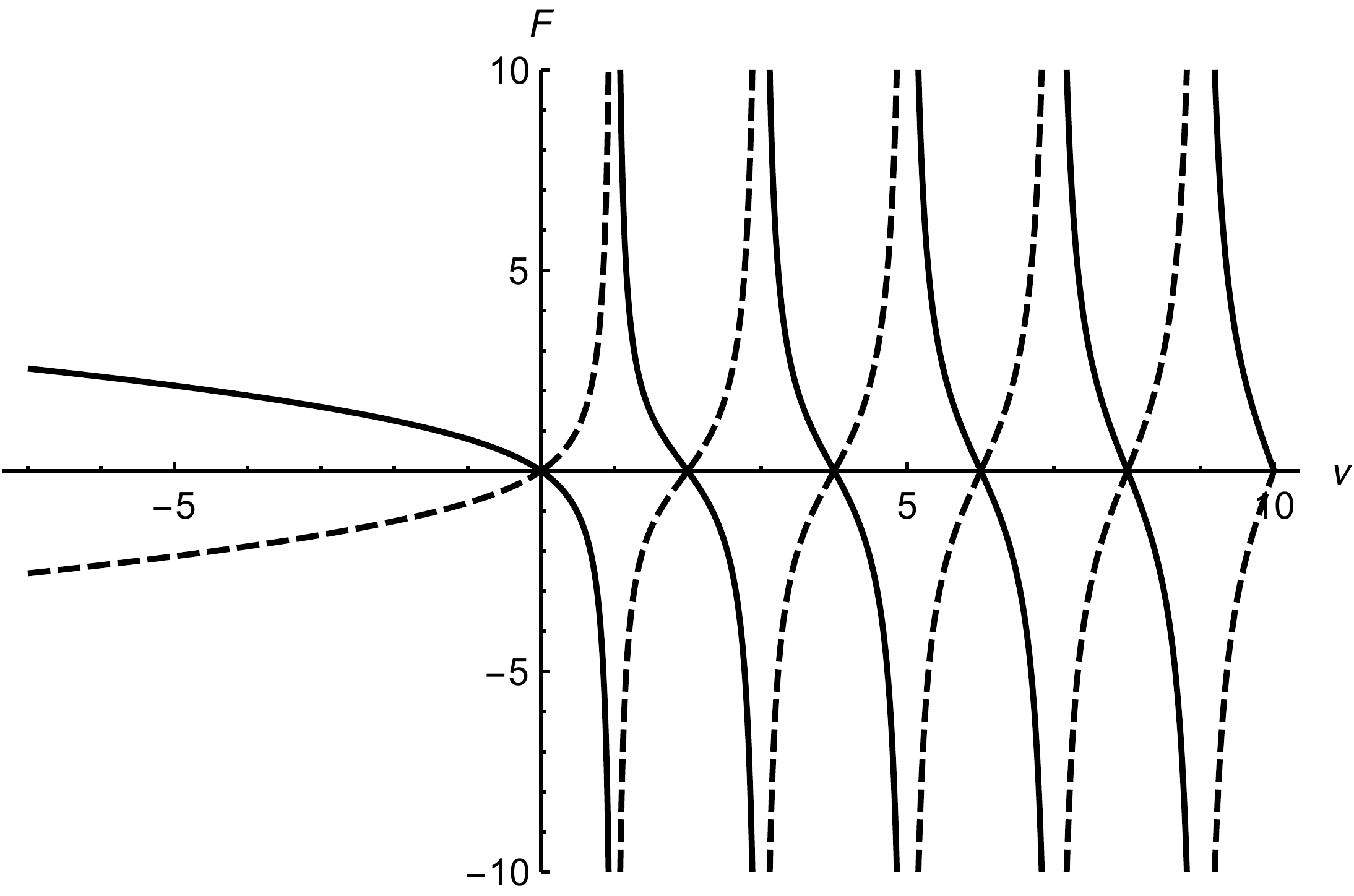}
 \caption{Functions $F(\nu)$ and $-F (\nu).$ The case $ s=1 $ of the symmetric harmonic oscillator.}
\end{figure}

\section{Wavefunctions}

The frequencies $\omega_+$ and $\omega_-$ are input parameters given by the springs. The corresponding eigenvalues $E_{\nu +} = \hbar \omega_+ (\nu_+ +\frac{1}{2})$ can be labeled by $n=0, 1, 2, \ldots,$ $\nu_n = \nu_{n+}$ in increasing order. (Note: numbers $\nu_+$  depend only on the frequency ratio $s.$) The number $n$ is also equal to the number of zeros of the corresponding eigenfunction. Eigenfunctions can be written using the Heaviside step function  $\theta(x)$
$$\psi(x)_{\nu_n} =\theta(x)(D_{\nu_{-n}}(0) D_{\nu_{+n}}(\sqrt{\omega_+}x))+ \theta(-x) (D_{\nu_{+n}}(0) D_{\nu_{-n}}(-\sqrt{\omega_-} x)).$$
  By dividing these eigenfunctions by the norm, we can introduce Fock's notation and we get elements of the orthonormal basis on space  $L^2(\R)$. 
$$\mket{s, n}  = \frac{\psi(x)_{\nu_n}}{\sqrt {\int_{- \infty}^{\infty}|\psi(x)_{\nu_n}|^2 dx}},$$ 

The expected value of the position of an asymmetric quantum harmonic oscillator in the state $ \mket {s, n} $ is $$ <x> = \mbra {s, n} x \mket {s, n}, $$ and it can be calculated numerically. 
An interesting effect occurs when the oscillator is in a state given by superposition of two eigenfuctions. For illustration, let the oscillator be in the  superposition  $$\mket{\psi}=\frac{1}{\sqrt{2}} (\mket{s,n}+\mket{s,k}).$$ 
 The time evolution of the mean value of the position is
$$\mbra{\psi} e^{i \mop H t} x e^{-i\mop H t} \mket{\psi} = \frac{1}{2} (\mbra{s,n} x \mket{s,n} + \mbra{s,k} x \mket{s,k}) + \mbra{s,n} x \mket{s,k} \, \cos(\omega_+ (\nu_n - \nu_k)t), $$
it oscillates around the point $ \frac{1}{2} (\mbra{s,n} x \mket{s,n} + \mbra{s,k} x \mket{s,k}) $ with the frequency $\omega_+ (\nu_n - \nu_k).$ It is  because $\mbra{s,n} x \mket{s,k}= \overline{\mbra{s,n} x \mket{s,k}}$ is non--zero. The frequency and the "amplitude" of the  asymmetric oscillator "quantum beats" depend on the difference of eigenvalues and  on the frequency $\omega_+.$ Some examples are given in Table 2. 
 
\section{Examples} 

Let us illustrate as an example the case $s=\sqrt{11}.$ The $\nu_+$ can be found, for example, using Mathematica program. We use command
\textsf{NSolve[Gamma[1/2 (1 - (x s + s/2 - 1/2))]/
    Gamma[-(1/2) (x s + s/2 - 1/2)] + (Sqrt[s] Gamma[(1 - x)/2])/
    Gamma[-(x/2)] == 0, \&\& $-1 < x < 11$}, 
    and it gives all $\nu_+$ between -1 and 11 with 16 digits precision.

  Taking  $ \nu_{+23} =  10.3881$ as an example and $\omega_+=2,$ the wave function is shown in Fig.4,
   and the classical and the quantum probability distributions are in Fig.5.
     \begin{figure}[h]
 \includegraphics[scale=0.6]{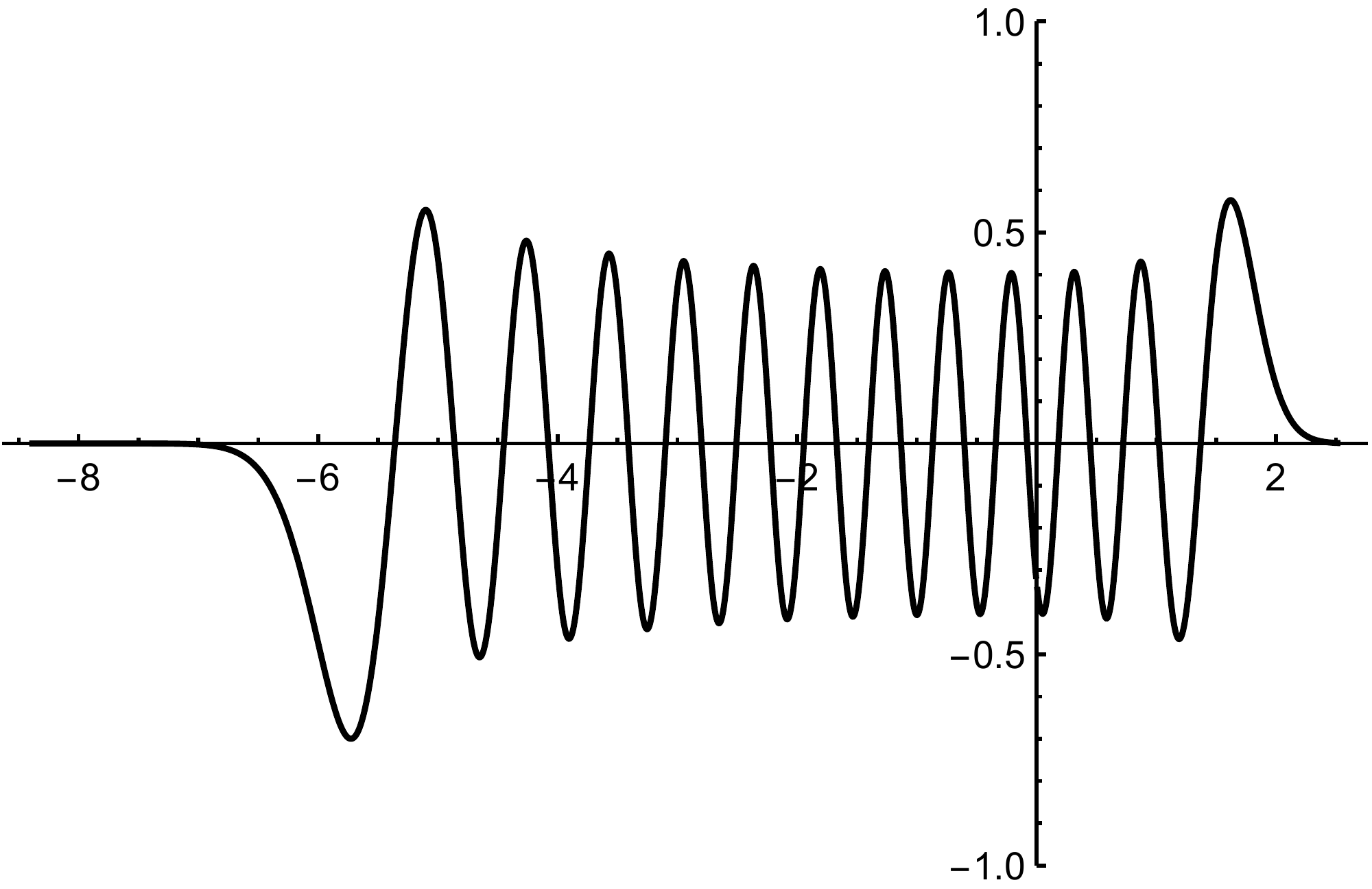}
 \caption{The wave function}
\end{figure}   
  \begin{figure}[h]
  \includegraphics[scale=0.6]{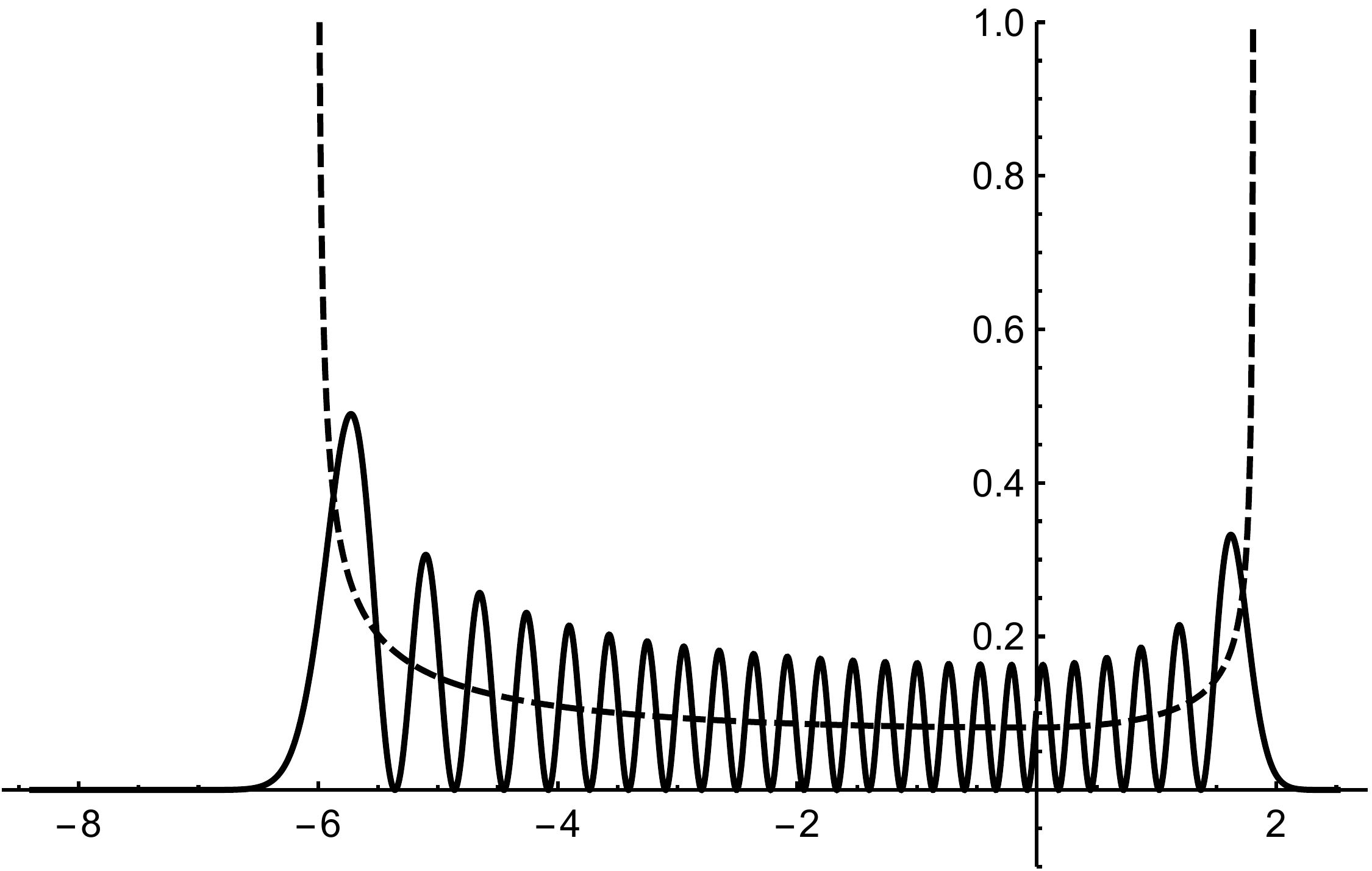}
  \caption{The probability distributions - classical and quantum}
\end{figure}
 \begin{table}[h]
 
\caption{The values of $\mbra{\sqrt{5},i} x \mket{\sqrt{5},j},\,i,j= 0,1,\ldots$}

\begin{tabular}{|l l l l l l l l |}

 -0.1321 & -0.4213 & -0.0536 & 0.0161 & -0.0040 & -0.0001 & -0.0009 &  0.0006 \\
  -0.4213 &  -0.2530 &  0.5851 & -0.0696 &  0.0180 & -0.0036 & 0.0005 & -0.0010 \\
  -0.0536 &  0.5851 &  -0.3303 &  -0.7189 & 0.0791 & -0.0198 & -0.0041 & -0.0003\\
 0.0161 & -0.0696 & -0.7189 &  -0.3858 &  -0.8320 &  0.0903 &  0.0230 &  -0.0049 \\
-0.0041 & 0.0180 & 0.0791 & -0.8320 & -0.4380 & -0.9286 & -0.1010 &  0.0253 \\
 -0.0001 & -0.0036 & -0.0198 & 0.0903 &  0.9286 & -0.4862 &  1.0171 & -0.1090  \\
  -0.0010 &  0.0005 &  -0.0041 &  0.0230 & -0.1010 &  1.0171 &  -0.5272 &  -1.0995\\
0.0006 & 0.0010 & -0.0003 &  -0.0049 & 0.0253 & -0.1090 & -1.0995 &  -0.5562 \\

 \end{tabular}
\end{table}   

\section*{Acknowledgement}
All authors acknowledge the financial support from RVO14000 and "Centre for Advanced Applied Sciences", Registry No.CZ.02.1.01/0.0/0.0/16\_019/0000778, supported by the Operational Programme Research, Development and Education, co-financed by the European Structural and Investment Funds and the state budget of the Czech Republic.

The author (G.C.) thanks V. Poto\v cek and J. Tolar for helpfull discussion.

 \end{document}